\documentclass[12pt]{article}
\usepackage{a4,latexsym,graphicx,times,color}
\usepackage[numbers]{natbib}

\graphicspath{{Figures/}}

\newcommand{\fg}[1]{\textcolor{blue}{ {#1}}} 

\newcommand{\showfigures}[1]{#1} 

\begin{document}

\title{Quasicrystalline three-dimensional  foams}
\author{{\bf \Large S.J. Cox$^1$, F. Graner$^2$, R. Mosseri$^3$, J.-F. Sadoc$^4$}\\
$^1$  Department of Mathematics, Aberystwyth University, SY23 3BZ, UK\\
$^2$ Laboratoire Mati\`ere et Syst\`emes Complexes, Universit\'e Denis Diderot - Paris 7, \\ CNRS UMR 7057, 10 rue A. Domon et L. Duquet, F-75205 Paris Cedex 13, France\\
$^3$ LPTMC,  UPMC, CNRS UMR 7600, Sorbonne Universit{\'e}s,\\ 4 place Jussieu, F-75005 Paris, France\\
$^4$ Laboratoire de Physique des Solides, CNRS-UMR 8502, B{\^a}t. 510, \\Universit{\'e} Paris-sud Paris-Saclay, F 91405 Orsay cedex, France.
}
\date{\today}

\maketitle

\begin{abstract}
We present a numerical study of quasiperiodic foams, in which the bubbles are generated as duals of quasiperiodic Frank-Kasper phases.  These foams are investigated as potential candidates to the celebrated Kelvin problem for the partition of three-dimensional space with equal volume bubbles and minimal surface area. Interestingly, one of the computed structures falls close (but still slightly above) the best known Weaire-Phelan periodic candidate. This gives additional clues to understanding the main geometrical ingredients driving the Kelvin problem.
\end{abstract}

\section{Introduction}

Aqueous foams are space-filling packings of bubbles, with soap film surfaces and configurations governed by the celebrated Plateau's laws. The physics of foams covers a large variety of systems and associated properties \cite{Cantat2013}. In parallel, foams prove interesting in solving and/or illustrating well-posed geometrical questions related to minimal surfaces. A paradigmatic example is the Kelvin problem \cite{kelvin87,weaire94}: {\it What is the least surface area of equal-volume objects which partition space? What is the global energy minimum of a foam in which the bubbles all have exactly the same volume?} These two formulations, one in mathematics and one in physics, are equivalent because the energy of a foam is proportional to the surface area of its soap films. The problem is difficult because there are many candidates and they all have comparable surface areas.

Kelvin himself conjectured a periodic candidate, where each (identical) bubble has the shape of a truncated octahedron with slightly curved surfaces
 \cite{kelvin87,weaire94}. A better periodic candidate was found a century later by Weaire and Phelan \cite{WeaireP94b}, who studied a periodic arrangement of two topologically-distinct types of bubbles with respectively 12 and 14 faces. The bubble centroids are located at the vertices of the well known A15 Frank-Kasper (F-K) phase \cite{frankkasper}. F-K phases are polytetrahedral periodic packings, which can be easily dualized, leading to plausible skeletons for foam structures.  As a consequence, many other F-K phases have been (unsuccessfully) tested with respect to the Kelvin problem (e.g. \cite{kraynikrvs02}), leaving the Weaire-Phelan candidate still unbeaten. 

F-K phases play an important role in a rather different area of material science, that of quasicrystalline materials. Indeed, they often appear nearby in phase diagrams, and display quite similar local atomic order (especially with icosahedral patterns).  It is possible to generalize the concept of a F-K phase to non-periodic structures, and for instance to define a quasicrystalline  F-K phase by atomic decoration on top of a quasiperiodic tiling \cite{sadocmosseri16}. The aim of the present paper is to analyze the related quasiperiodic foams with respect to the Kelvin problem.

Section 2 gives an introduction to foams, Plateau's laws, and the Kelvin and Weaire-Phelan candidates. Section 3 describes the quasicrystalline F-K structures and their dual bubble skeletons. Section 4 presents our numerical results on quasiperiodic foams, compared to previously obtained results for periodic foams. Section 5 summarises our approach, discusses the results and opens perspectives.

\section{From Kelvin to Weaire-Phelan}

\subsection{Foams}

Kelvin's initial motivation \cite{kelvin87,weaire94} was to look for a possible structure for the aether that permeated space. He required a structure through which light could propagate like sound does in an elastic solid medium. The puzzle was that light admits only two transverse polarizations, while sound also admits a third, longitudinal one. Kelvin looked for a low-density material, with a high ratio of its bulk to shear elastic moduli, and an elegant structure close to perfection. Thus he looked for a foam. 

Nowadays, the interest in the aether has faded, but not the interest in foams, which have numerous industrial properties in addition to being aesthetic and useful for scientists \cite{Cantat2013}.  In addition, this unsolved problem specifically attracts mathematicians and physicists to work together, because the simplicity of the question contrasts with the difficulty of addressing it \cite{ini}, and even draws the attention of architects seeking efficient partitions of space and optimal use of material.  
Of course we do not expect a foam to spontaneously find the global minimum of surface area among the many, many local minima. But the parallels between aqueous foams and least-area partitions of space provide a rich source of cross-fertilisation \cite{WeaireH99}. 

An equilibrium dry aqueous foam is a collection of bubbles that fit together without gaps or overlap. The shape of each interface between two bubbles, i.e. each soap film, is governed by surface tension, which acts to reduce the surface area of each film to a minimum. The volume of gas in each bubble can be kept fixed (at least over short to intermediate time-scales of seconds to minutes) and, under controlled conditions (see e.g. \cite{hohlerclca08}), foams with bubbles of equal volume can be produced. Hence such a foam is an area-minimizing space-filling structure, and each realisation can be thought of as a candidate structure that meets the conditions of the Kelvin problem.

From foams comes the solid background to build upon, called Plateau's laws \cite{Cantat2013,WeaireH99,plateau73}. In any least-area partition the interfaces must meet three-fold along edges (at equal angles of $120^\circ$), and the edges along which the interfaces meet themselves meet four-fold at vertices with the tetrahedral angle, $\theta_t= \arccos( -1/3) \approx  109.5^{\circ} \approx 1.91$ rad. That these laws are a consequence of surface area minimization was proven by Taylor \cite{taylor76}, building on work by Almgren \cite{almgrent76, morgan4th}.
A purely topological consequence of the fact that edges are three-fold and vertices are four-fold which in this case arises from Plateau's laws, but which can also be observed in systems which do not obey them) is that, on each  individual bubble, the  number of faces $F$ and the average number of edges per face $\langle e_F \rangle$ are linked through  \cite{Cantat2013} $ (6 -\langle e_F \rangle) F = 12$.

Further, the Young-Laplace law relates the bubble pressures to the (mean) curvature of each interface. If two bubbles share a face, this face has a constant mean curvature, while separately its radii of curvature can vary, and its Gaussian curvature too. If two bubbles which share a face have the same pressure, this face has uniformly zero mean curvature, and hence at each point its radii of curvature are opposed, although both can be non-zero like the Gaussian curvature.

\subsection{Periodic structures}

The Kelvin problem has a simpler counterpart in 2D:
what is the minimal perimeter of equal-area objects which tile the plane? 
The solution, a tiling of regular hexagons, was known to the Romans from their observation of beehives, but its rigorous,  computer-aided proof is recent  \cite{hales01}. Regular hexagons have the same shape, the same pressure, flat sides, obey Plateau's laws, and tile the plane. Their perimeter is only a few percent above that of the circle, which is the exact theoretical lower bound for any single object in general, but which does not tile the plane.

Back in 3D, there exists a simpler variant of the Kelvin problem: what is the minimum surface area of a bubble which tiles space by periodically repeating itself?
There is no shape which transposes to 3D all the properties of 2D regular hexagons.  
The main property one is tempted to generalize is that 2D regular hexagons have identical flat faces. Can one conceive a 3D bubble with exactly identical flat faces? If one blindly applies Plateau's laws, one finds that such a bubble would necessarily have  $e_F = 2\pi/(\pi - \theta_t) \approx  5.104$ edges per face and $\approx$13.39 faces\cite{isenberg92,avron92,cox03}. 
Note that these considerations, derived here for bubbles obeying Plateau's laws, correspond to one of the two ideal solutions for three-dimensional packings -- the trigonometric solution rather than the algebraic one -- discussed by Coxeter \cite{coxeter58,coxeter61}, both of which he called ``statistical honeycombs".

Such ideal bubble packing is of course impossible to realize in practice. Still, since we can calculate analytically the surface area of a regular bubble obeying Plateau's laws with any number of faces   \cite{Hilgenfeldt2004}, interpolation indicates that this hypothetical, so-called ``ideal" bubble   would have a surface area of $\approx$5.254  (hereafter, surface areas are given with respect to bubbles of unit volume). It is reasonable to conjecture that no actual bubble could beat this value, and hereafter we use it as a reference for comparison with various candidates. Note that this lower bound for bubbles obeying Plateau's laws and tiling the space is only a few percent above  the exact theoretical lower bound for any single object in general, which is that of the sphere ($\approx$4.836).

\begin{figure}
\centerline{
(a)
\showfigures{\includegraphics[width=0.4\textwidth]{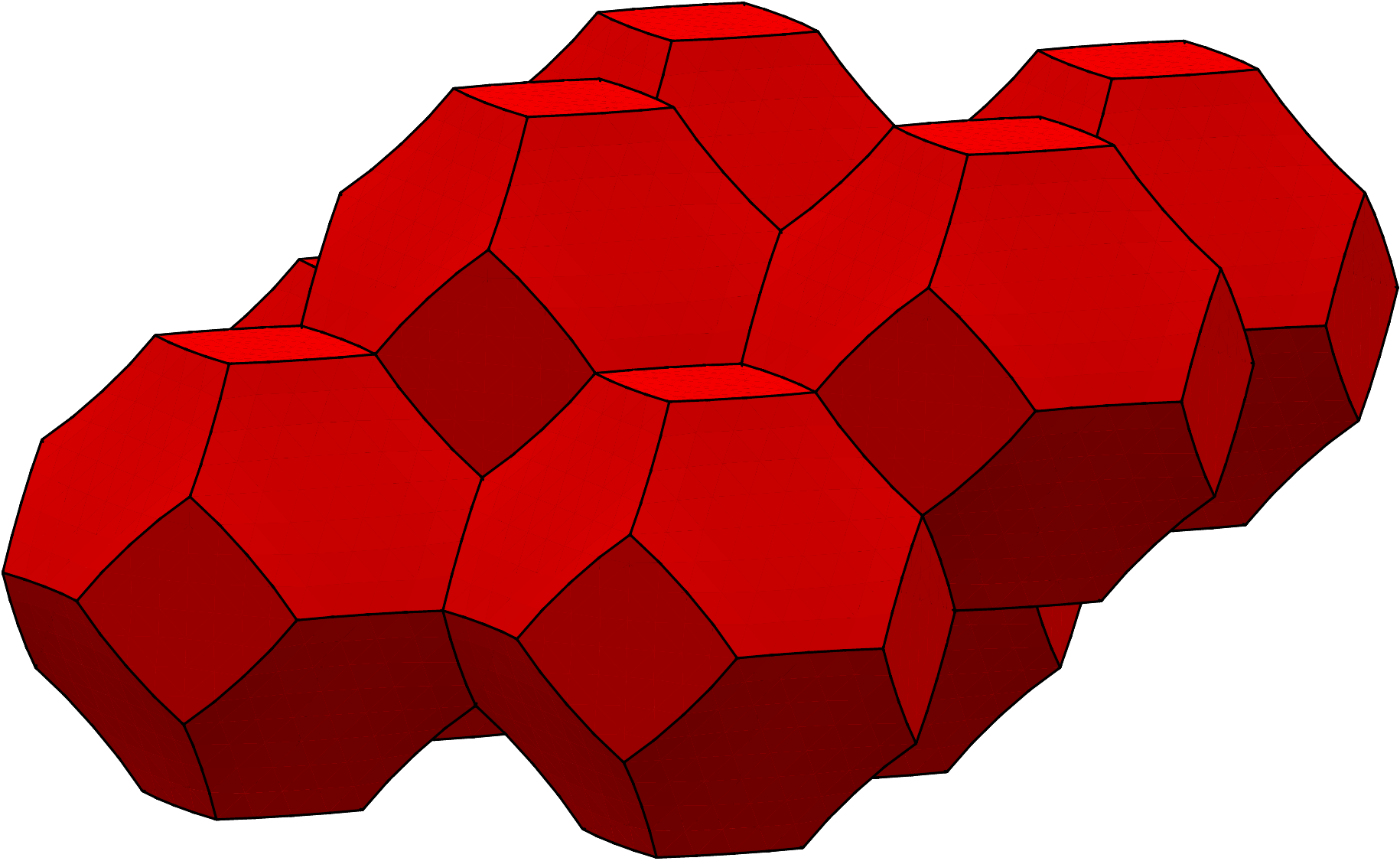} }
(b)
\showfigures{\includegraphics[width=0.35\textwidth]{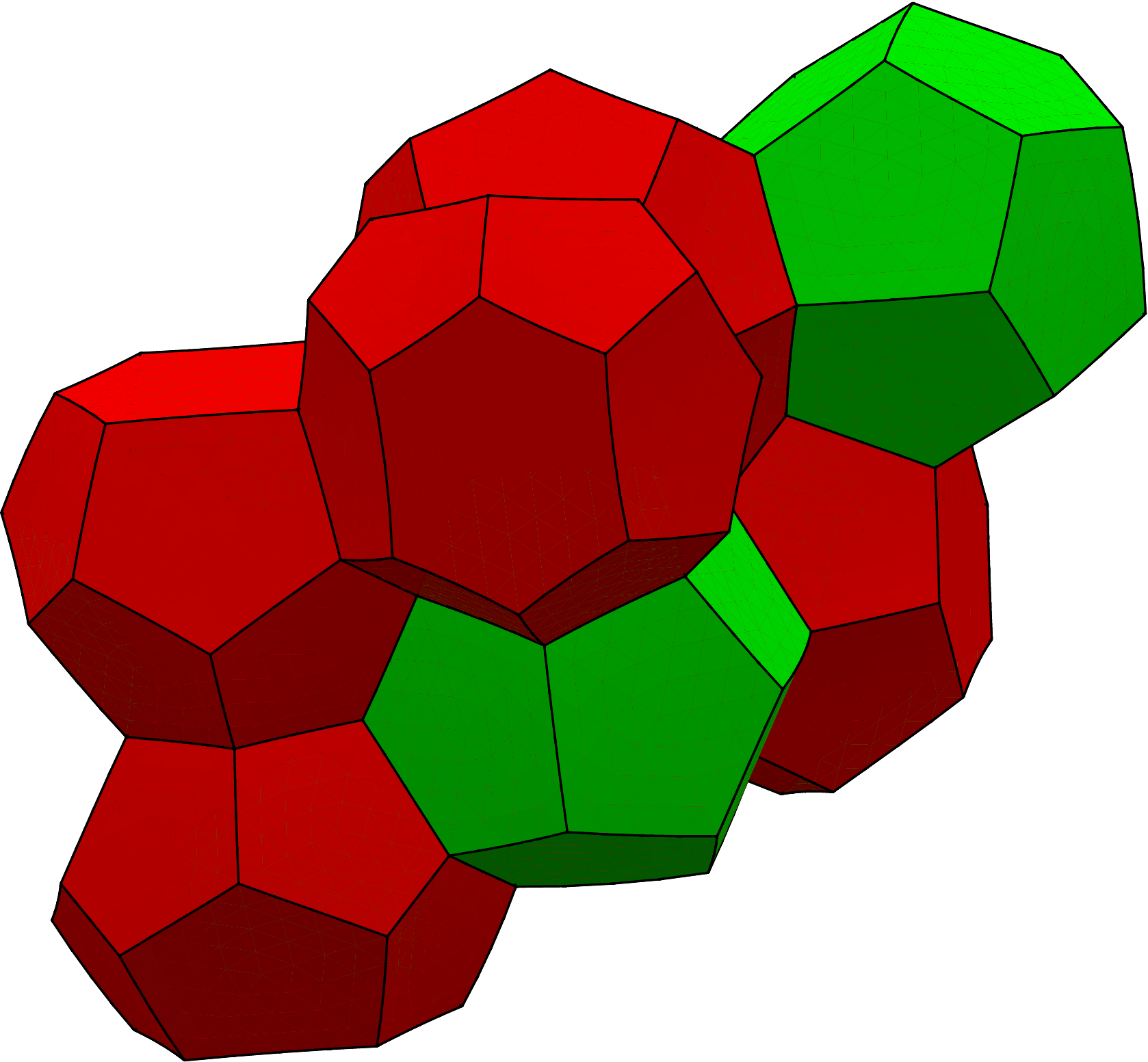} }
}
\caption{Periodic structures. (a) Eight repeated bubbles from the Kelvin structure. (b) The eight-bubble unit cell of the Weaire-Phelan structure.}
\label{fig:kelvin_wp}
\end{figure}

For actually realisable bubbles, one looks for a unit cell which periodically repeats itself, as in a monatomic crystal\fg{,} with faces which have  everywhere two opposed radii of curvature, and obeys Plateau's laws. 
The candidate proposed by Kelvin \cite{kelvin87} is known as the Kelvin structure, the truncated octahedron or the tetrakaidecahedron. It has
eight hexagonal faces and six square faces, hence 14 faces and an average of 5.14 edges per face, close to that of the hypothetical ideal bubble. It requires ``delicate'' \cite{kelvin87} curvature of its faces to satisfy the conditions of space-filling and minimal surface area (fig. \ref{fig:kelvin_wp}(a)). Since all the bubbles have exactly the same shape, they have the same pressure and consequently every face has zero mean curvature.
The surface area of Kelvin's  truncated octahedron  is $\approx$5.306.

Coming back to the full Kelvin problem, there is no requirement that all bubbles should have the same shape. Bubbles with different shapes, but obeying Plateau's laws, have very similar surface areas  \cite{Hilgenfeldt2004}, and that makes the Kelvin problem difficult.  Formally, it remains unsolved: there is no proof that a particular structure is the least-area way to partition space into equal-volume bubbles.
In general, there is no analytic method to estimate a structure's surface area. One usually needs to determine theoretically the positions of sites, then find their Voronoi structure, and simulate with a high precision the relaxation of its surface towards a local minimum, usually using  the Surface Evolver \cite{brakke92}. 
What we now know for sure, is that Kelvin's truncated octahedron is not the global minimum. 

In fact, it has been beaten, and a better candidate was found a hundred years after Kelvin.
The guiding idea was that the hypothetical ``ideal" flat-faced bubble was probably the ultimate limit towards which we should tend; any deviation of $e_F$ from 5.1 edges per face (or, equivalently, of $F$ from 13.39 faces per bubble) would imply the existence of curvature, and thus probably a cost in terms of surface area. In 1994, Weaire and Phelan looked for bubbles such that {\it each} face should have close to  5.1 edges (that is, not just on {\it average}, like the truncated octahedron, which has no pentagonal faces). As a consequence, the strategy  to further lower the surface area was to  introduce as many five-sided faces as possible, while not adding too many four- or six-sided faces (and neither three nor seven-sided ones). As bubbles with only pentagonal faces are dodecahedra and are unable to tessellate space on their own, it is necessary to mix bubbles of at least two different shapes (but, for this problem, with the same volume). Several candidates, inspired by known polyatomic crystalline structures, were numerically simulated using the Surface Evolver software.  
The best candidate, thereafter known as the Weaire-Phelan (W-P) structure \cite{WeaireP94b}   (fig. \ref{fig:kelvin_wp}(b)),
 is a relaxed form of the dual to the F-K A15 (or $\beta$-tungsten) structure, consisting of two types of bubbles with 12 faces (pentagonal dodecahedra) and 14 faces (Goldberg barrels). 
It has a surface area of $\approx$5.288. 
Many further periodic structures were investigated by Weaire and Phelan and others without finding one with lower surface area than W-P.

%
%

\subsection{Quasiperiodic structures}

We have tried to extend the Weaire-Phelan approach by keeping the idea of having bubbles with different shapes and topologies, but relaxing the constraint of having a periodic structure, which is not a prerequisite in the Kelvin problem. A natural way to do that is to investigate quasiperiodic bubble arrangements, and more specifically Frank-Kasper quasicrystalline structures.

The experimental discovery of icosahedral quasicrystals \cite{shechtmann84} represents a major breakthrough for materials science, opening up a wider range of stable atomic structures in the solid state, between periodic crystals and disordered systems. 
Quasicrystals allow for interesting long and short-range orientational order, especially five-fold symmetries forbidden to real crystals; starting from tetrahedrally-close-packed quasicrystals, their dual structures (which will form the bubble skeleton) will have a high density of pentagons. 

Compared to periodic structures, quasicrystals have an important drawback: to relax their Voronoi structure and determine their surface area  it should be necessary to simulate an infinite number of sites. 
We thus ask the following questions :

- Can we theoretically select quasicrystals in which the distribution of edge numbers is peaked around 5.1 and, if possible, the distribution of the number of faces is peaked around 13.39?

- Can we turn these theoretical structures into finite-size simulations of foams, with sufficient precision to discriminate between the surface area values of different structures, and with a method to validate this precision?

- Does a structure's surface area correlate with the difference between its average number of edges per face and the value 5.1? And/or with the difference between the average number of faces per bubble and the value 13.39?

\section{Construction of quasiperiodic foams}

 \subsection{Frank-Kasper phases}

\begin{figure}[tc]
\showfigures{\includegraphics [width=14cm] {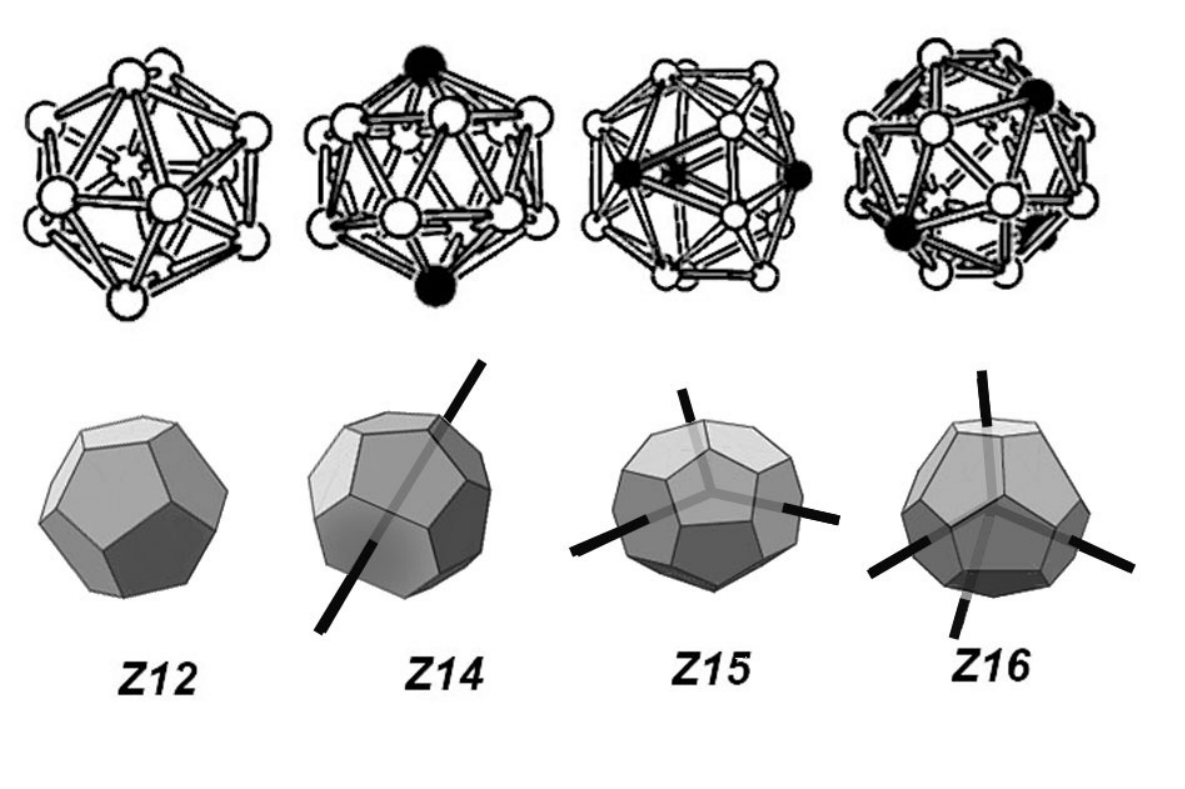}}
\caption{\label{fig:FK}Frank-Kasper structures. The four main Frank-Kasper coordination polyhedra (top row)  $Z_{12}$,  $Z_{14}$,
 $Z_{15}$ and $Z_{16}$, represented also by their Voronoi domains (lower row). Frank-Kasper lines (or disclination lines), shown as black lines, start from the centre of the coordination polyhedra and run through the black sites, or dually from the Voronoi domain centres, and go through hexagonal faces, defining a ``major skeleton" in the structure.}

\end{figure}

The stable quasicrystalline metallic alloys occupy a narrow range in their respective phase diagrams, stimulating an analysis of their structure and the main ingredients responsible for their stability, together with that of the neighbouring crystalline phases, generically large unit cell crystals among which can be found the Frank-Kasper-like phases \cite{frankkasper}. It is standard in the latter case to analyze  the structure in terms of local atomic environments, the canonical $Z_p$ coordination cells (see fig.~\ref{fig:FK}), with $p$ the coordination number, the case $p=12$ corresponding to a local icosahedral environment.

Many F-K structures can be generated with an atomic decoration procedure applied to a plane tiling made of triangles and/or squares, building simple atomic layers. These rules are not limited to the periodic case; as an example, quasicrystalline F-K phases can be generated from a dodecagonal  quasiperiodic tiling \cite{sadocmosseri16}. Related structures have been invoked to model dodecagonal quasicrystals found in metallic alloys \cite{ishimasa,chen} and, more recently, in dendrimeric supramolecular liquid crystals \cite{zeng,ungar}. The latter discoveries are particularly interesting because they show that metallic bonding is not a prerequisite for stable quasiperiodic order. 
Further, F-K phases were also  observed in micellar structures \cite{hajiw}, in which amphiphilic molecules form micelles in water or, in an inverse form, separating drops of water with films that are minimal surfaces.

As noted above, the conjectured best W-P candidate is dual to the $A15$ F-K structure, while the original Kelvin candidate is dual to the BCC packing. Frank-Kasper structures are tetrahedrally-close-packed (TCP) structures: the atomic positions decompose three-dimensional Euclidean space into tetrahedral unit cells which are not far from being regular. It is well known that regular tetrahedra
 cannot fill space perfectly, leading to geometrical frustration \cite{sadoc1981,sadocmosseribook}. 
The  tetrahedron's dihedral angle is not an integer submultiple of $2\pi$, although it  is close to $3\pi/5 \approx 1.885$ and hence $\pi -\theta_t$ is close to $2\pi/5$. One therefore expects to find a large proportion of edges sharing five tetrahedra;  whenever all edges through a site are of this type, the local order is icosahedral, and the site corresponds to a F-K canonical $Z_{12}$ one (fig.~\ref{fig:FK}). Other coordination polyhedra $Z_{p}$, mostly with  $p=14, 15$ or $16$, are found in real TCP structures, with an average, structure dependent, coordination number around $p=13.4$. 

Beside the $Z_{p}$ distribution, another important parameter concerns the F-K major skeleton \cite{frankkasper}, later identified as a disclination network \cite{sadocmosseri1982,sadoc1983,nelson1983,sadocmosseri1984}, formed by edges sharing six tetrahedra, with two types of sites: (i) edge sites, made of $Z_{14}$ polyhedra,  threaded by the edge through opposite points with hexagonal symmetry and (ii) vertex sites, where disclinated edges meet in threes at an angle of $2\pi/3$ on $Z_{15}$ sites, or by four at an angle $\theta_t$ on $Z_{16}$ sites. These disclination lines cannot be interrupted in the structure: they run throughout the volume, and can connect to other lines.

Among the large set of F-K structures, A15 is a particular, extremal, case: it contains only $Z_{12}$ and $Z_{14}$ sites, with a rather high average coordination number $p= 13.5$, and a disclination network formed by periodic disconnected straight lines running in the three perpendicular directions. To check the importance of these peculiarities, it was therefore tempting to compute the associated value of surface area for more generic F-K phases; this was done already for several such structures \cite{kraynikrvs02,kusners96}, always showing larger values than for W-P.

Our aim here is to widen the analysis of TCP duals by considering two families of quasiperiodic F-K foams, allowing us to compare the obtained surface areas with those of formerly studied periodic F-K structures, and order these new ones according to the relative occurrence of 15-fold and 16-fold coordinated vertices.

%
%

\subsection{Quasiperiodic Frank-Kasper phases}
\label{sec:FK}

%
\begin{figure}[tbp]
\showfigures{\includegraphics [width=14cm] {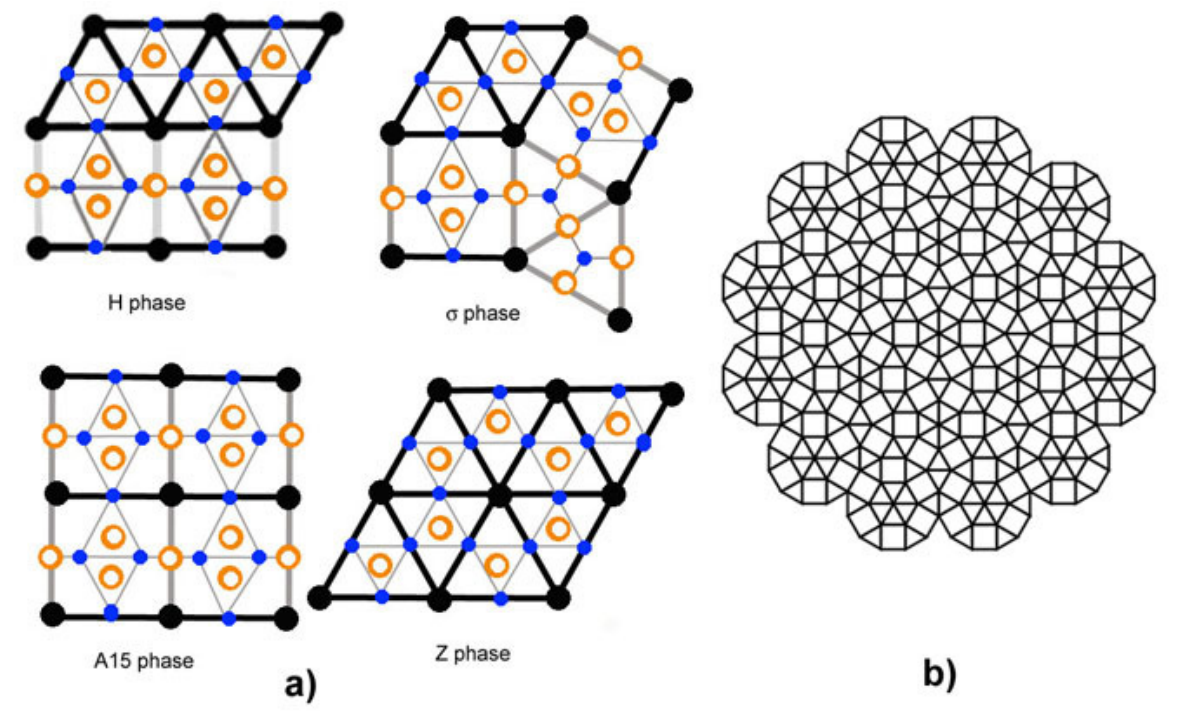} }
\caption{Layered Frank-Kasper phases and triangle-square tilings.
(a) Some well-known F-K phases, and their underlying tilings by squares and/or triangles; atomic positions at height $h=1/4$ or $3/4$ are shown in black, those at height $h=1/2$ in blue, and those at $h=0$ as open circles.
(b) A piece of an (undecorated) dodecagonal quasiperiodic tiling, which underlies the F-K quasicrystalline phases discussed in the text.
}
\label{fig:squaretrianglefk}
\end{figure}
%

We now briefly present two families of recently-derived quasiperiodic structures with dodecagonal symmetry, described with more details in ref. \cite{sadocmosseri16} (see also ref. \cite{mosserisadoccras}). 
They belong to the large set of ``layered" F-K structures, in which atomic positions can be gathered into simple planes defined relative to an underlying tiling template made of squares and/or triangles, see for instance \cite{sullivan,sikiric}. Once the tiling is given, four atomic layers are generated, say at vertical coordinates $0,1/4,1/2$ and $3/4$. Two parallel copies of the tiling itself lie at heights $1/4$ and $3/4$, with atomic positions at the vertices.
Some well-known F-K phases are presented in fig. \ref{fig:squaretrianglefk}a, with their associated underlying tiling. Also shown, in fig. \ref{fig:squaretrianglefk}b, is a piece of a dodecagonal quasiperiodic tiling which will serve as a template in the quasicrystalline case.

\begin{figure}[tbp]
\showfigures{\includegraphics [width=14cm] {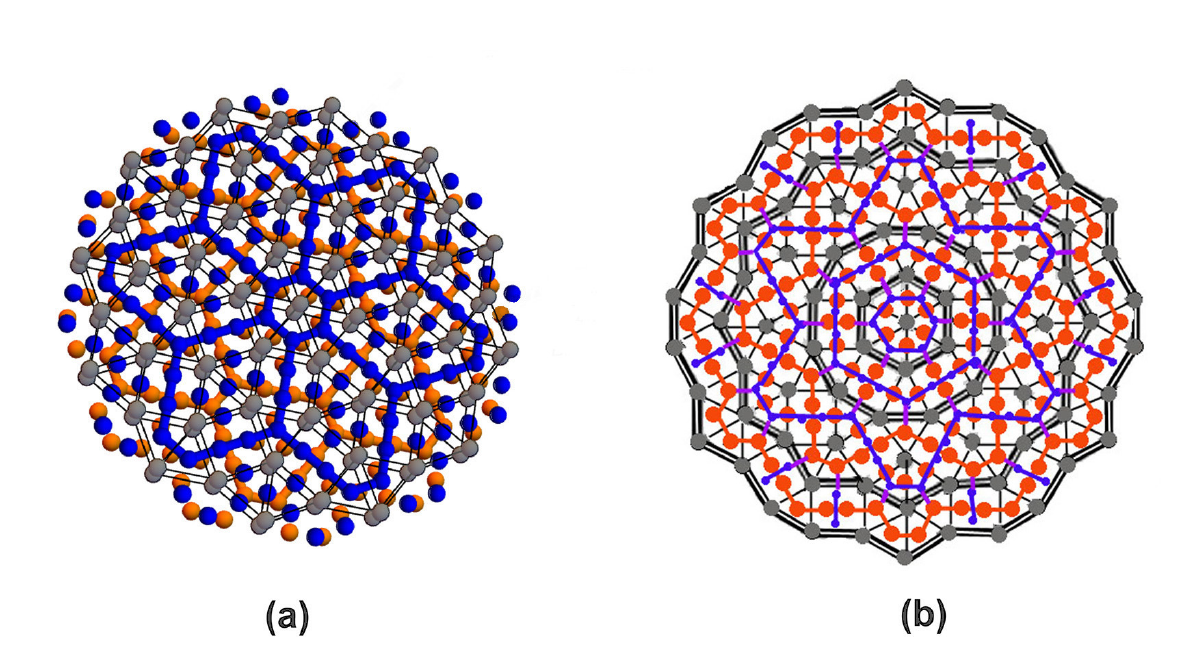} }

\caption{
Quasicrystalline dodecagonal Frank-Kasper phases. (a) Family A, denoted DQ-$\rm{A}$, a structure with only $Z_{12}$, $Z_{14}$ and $Z_{15}$ sites. The disclination network is represented inside two layers, respectively in red and blue (colour on-line). There are also disclinations (not drawn here) orthogonal to the layers, connecting grey sites located at the vertices of the triangle-square tiling. 
(b) Family B,  including $Z_{16}$ sites. The decoration requires drawing annular strips, delimited by double edges, where the disclination network connects different layers. We show here an example, denoted DQ-$\rm{B}_1$, with a dense array of concentric strips. For  clarity the drawing does not display the $Z_{12}$ sites belonging to layers at height $0$ and $1/2$, and instead focuses on the disclination networks with $Z_{14}$ and $Z_{15}$ sites.   Grey sites (located at heights $1/4$ and $3/4$) are either $Z_{12}$ sites on double edges or $Z_{14}$ sites on simple edges. Disclination networks are drawn in red (respectively in blue), colour on-line, for the layer at height $0$ (respectively $1/2$). Disclination segments crossing double edges (and connecting two different layers) appear in purple (colour on-line).
}
\label{fig:quasifk}
\end{figure}

From the dodecagonal square-triangle tiling, we generate three different structures, belonging to two families $\mathrm{A}$ and $\mathrm{B}$, which we denote by DQ for ``dodecagonal quasicrystal":

(i) Family $\mathrm{A}$, with only $Z_{12}$, $Z_{14}$ and $Z_{15}$ sites, which are such that the disclination network is formed by planar networks (along the layers), or perpendicular to the layers (see fig \ref{fig:quasifk}a). Once the triangle-square tiling is given, the decoration is unique, and easy to construct automatically; we denote it DQ-$\rm{A}$. The average coordination number for this quasicrystal is $p \approx 13.464$ and we can extract an (unnormalized) composition, in terms of the coordination numbers: $Z_{12}^rZ_{14}^sZ_{15}^t$ with $r=3+\sqrt{3}$, $s=2+3\sqrt{3}$ and $t=2$. 

(ii) Family $\mathrm{B}$, containing variable numbers of $Z_{16}$ sites, as well as new $Z_{15}$ sites, both contributing to connect the disclination network between the different layers. These connections appear transversally to a subset of the underlying tiling edges, subject to certain constraints, depicted by ``double edges". Given the underlying tiling, there is a large number of possible double edge decorations.  For the present study, we have constructed two such structures: DQ-$\rm{B}_1$, corresponding to the example shown in fig \ref{fig:quasifk}b, with a dense array of concentric double edge circuits and $Z_{16}$ sites; and DQ-$\rm{B}_2$ with a less dense array of double edge circuits, few $Z_{16}$ sites, and mainly new connecting $Z_{15}$ sites. Since we used a manual procedure to generate these structures, for which the required generation time increases significantly with the number of bubbles, and since these structures appeared to have a higher surface area, we have not built structures with more than 1500 bubbles. Note that in addition to the disclination network connecting the different layers, there are still linear disclinations perpendicular to the layers, but with a lower density compared to DQ-$\rm{A}$.

%
%

\subsection{Simulation methods}

\begin{figure}
\centerline{
}
\includegraphics[width=14cm]{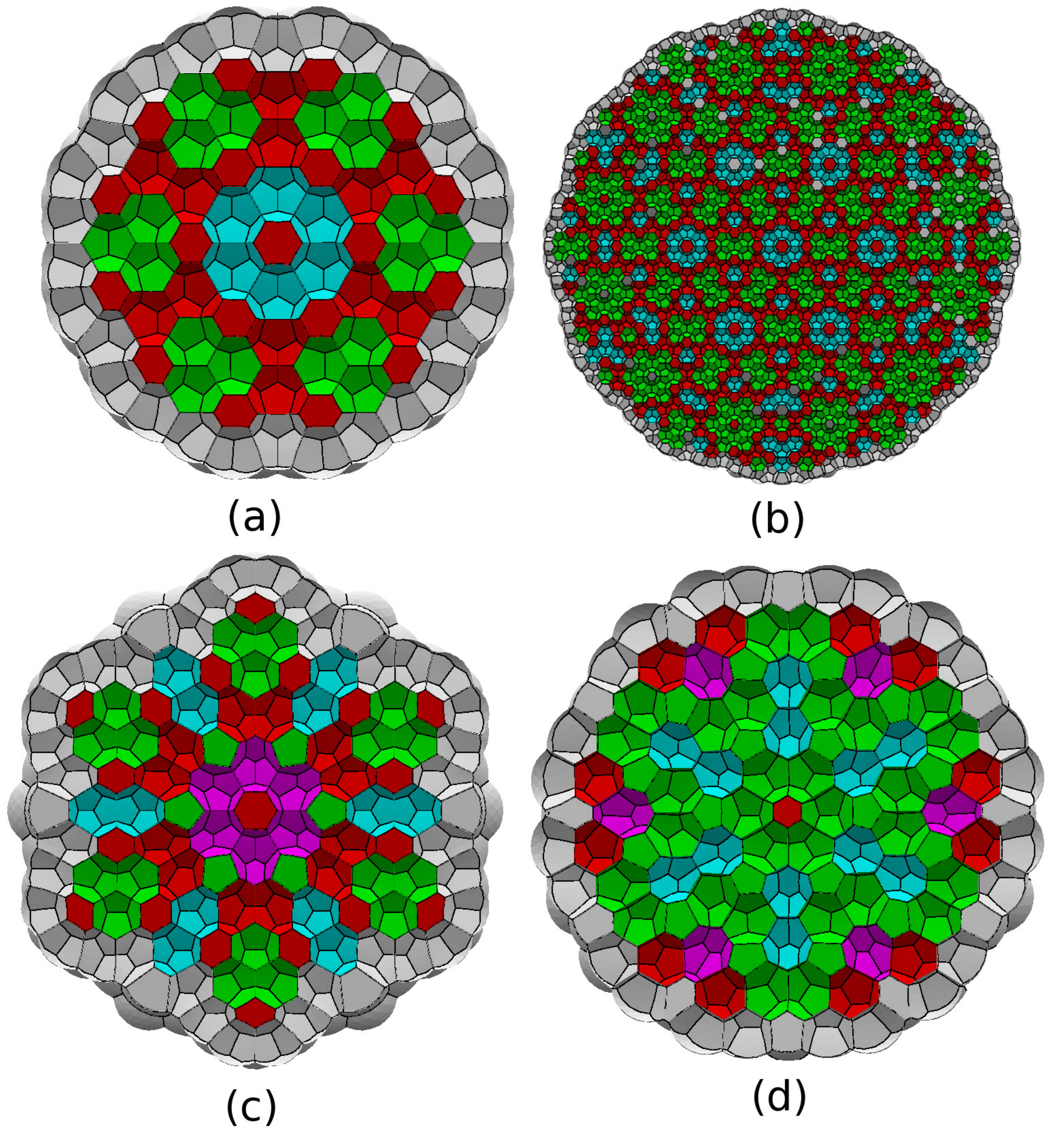}

\caption{
Examples of quasicrystalline foams. In each case the view is in the direction of periodicity, perpendicular to the $x$-$y$ plane, sliced at $h=\frac{1}{2}$. Bubbles that are not on the periphery of the cluster, and hence contribute to the surface area calculation, are coloured by their number of faces (colour on-line): $F$=12 - green, $F$=14 - red, $F$=15 - cyan, $F$=16 - magenta. (a) DQ-$\rm{A}$, small size. (b) A larger patch of DQ-$\rm{A}$, with about 2000 bubbles. (c) DQ-$\rm{B}_1$. (d) DQ-$\rm{B}_2$. 
}
\label{fig:clusterpic}
\end{figure}

We take the quasicrystalline F-K sites described above as the seed points for a Voronoi partition to generate  foam structures which are quasicrystalline in two directions. The structures are periodic in only one direction (perpendicular, by definition, to the $x$-$y$ plane) and must be truncated in the $x$-$y$ plane in some way to enable a finite-size simulation of a non-periodic structure. We choose to truncate each unconverged structure to lie within a circular disc of radius $R$ (fig. \ref{fig:clusterpic}), and to generate structures for increasing values of $R$, in the expectation that their surface areas will converge, for each type of quasicrystal, to a well-defined limit.

Sullivan's VCS software \cite{vcs} is used to generate the Voronoi cells from the seeds. The output from VCS is imported into Brakke's Surface Evolver, which is used to determine the surface area of each of our candidate structures (fig. \ref{fig:clusterpic}). In Surface Evolver the foam is made monodisperse (equal volume) by setting each bubble's target volume to be the average of the volumes given by the Voronoi partition; to allow for curved faces, each face is discretized with 10-20 triangles; and the total surface area of the foam is minimized, using both gradient descent and second derivative Hessian information, until it is accurate to four significant figures. The result of the truncation process is often a structure with a non-uniform outer edge (for example one Voronoi point might fall just inside the disc leading to a slightly protuberant bubble): we therefore do not include any bubbles on the periphery of the structure when calculating the surface area of the foam. 

To allow comparison of finite, quasicrystalline, structures with the fully periodic Kelvin and W-P structures, we also truncate the Kelvin and W-P foams in the $x$-$y$ plane. We therefore determine the minimum surface area of finite circular monodisperse clusters of these well-known foams, necessarily higher than the surface area of the corresponding periodic structures.

%
%

\section{Results}

\begin{figure}
\centerline{
(a)
\showfigures{\includegraphics[width=0.75\textwidth]{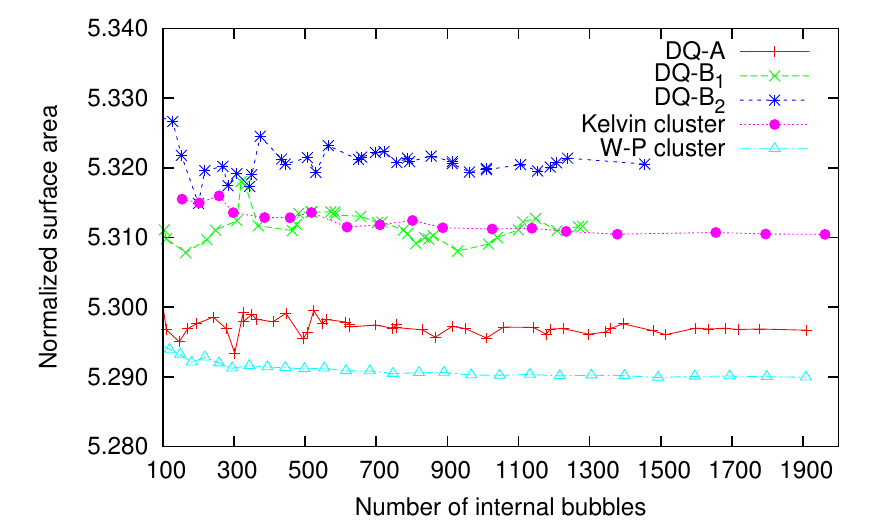} }
}
\centerline{
(b)
\showfigures{\includegraphics[width=0.75\textwidth]{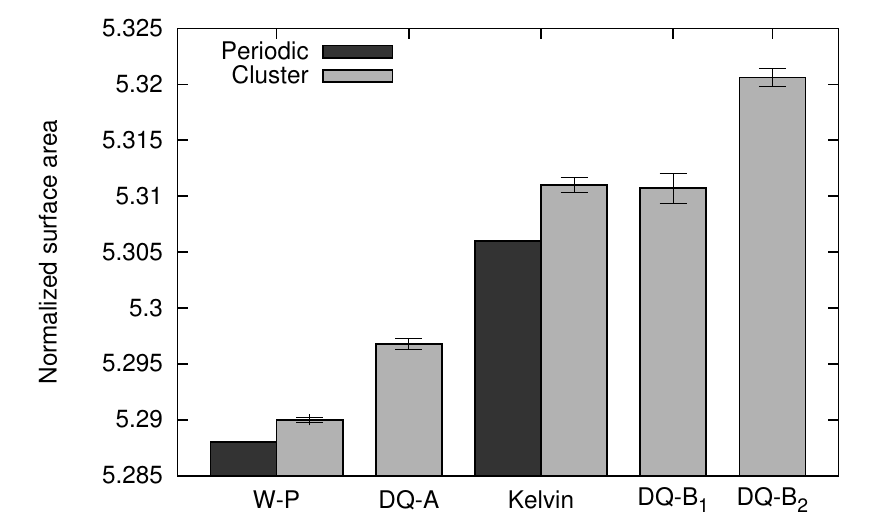} }
}
\caption{Comparison of areas. (a) Surface area, normalized by bubble volume, of all finite circular clusters considered, based on both periodic and quasiperiodic tilings. (b) The limiting value, measured as the average surface area ($\pm$std, plotted as bars) for $N\ge 700$, for the quasicrystal structures compared with the infinite Kelvin and Weaire-Phelan structures.}
\label{fig:surface_area}
\end{figure}

Our results are shown in fig. \ref{fig:surface_area}(a). Finite clusters consisting of Kelvin and W-P bubbles do tend towards the well-established limiting values of surface area given above, although the approach is slow -- we would require thousands more bubbles to reach it, beyond the available computational time -- and the fluctuations give an idea of the effect of our circular boundary condition. The quasicrystalline structures show even greater fluctuations, but still distinct behaviour. The DQ-$\rm{A}$ structure has by far the lowest surface area of the three, well below Kelvin and only slightly greater than W-P. DQ-$\rm{B}_1$ is broadly similar to finite Kelvin clusters in terms of surface area, although the structures are very different, and DQ-$\rm{B}_2$ is higher.

%
%

Fig. \ref{fig:surface_area}(b) summarizes the average surface area of each structure in comparison with the fully-periodic Kelvin and W-P structures. 
The difference in surface area between the finite circular clusters and the extended periodic tiling is as small as 0.1\% for Kelvin, and less than $0.05$\% for W-P. This suggests that our finite simulations accurately capture the limiting values for all structures.

\begin{figure}
\centerline{
\showfigures{\includegraphics[width=0.75\textwidth]{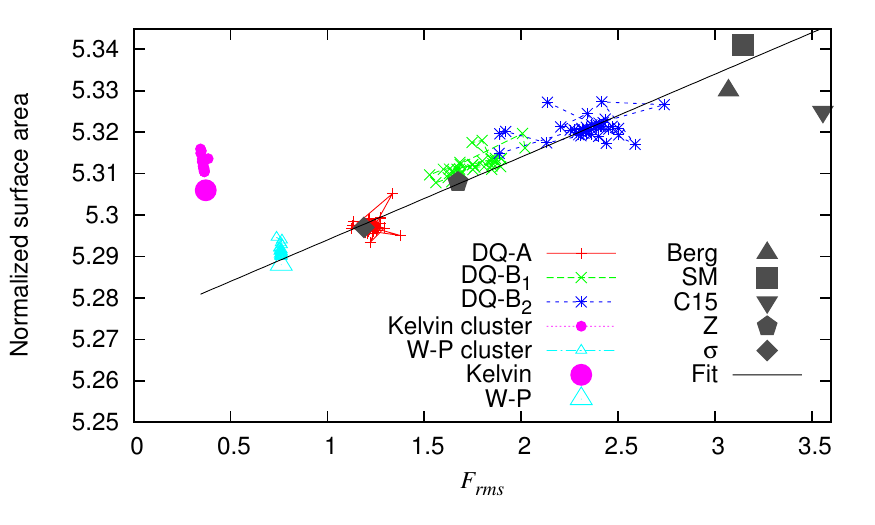} }
}
\caption{Area {\it vs.} topology. The normalized surface area of each simulated structure is shown against the root mean square deviation of its number $F$ of faces from that of the hypothetical ideal bubble, $\approx$13.39. The data from structures based on the F-K phases, including W-P, fall on a straight line ($5.274 + 0.02 F_{rms}$). Note also how close the known values for the infinite periodic foams, shown as larger points, are to the data for the finite circular clusters. For comparison, other structures discussed in the text are plotted with black symbols, and the origin of the ordinate axis is the surface area of the hypothetical ideal bubble.
}
\label{fig:S_vs_rmsF}
\end{figure}

We seek a topological parameter that succinctly describes the data. Each structure has an average number of faces per bubble $\langle F \rangle$ and average number of edges per face $\langle e_F \rangle$; the two are closely related, $\langle F \rangle = 12/(6-\langle e_F \rangle)$ and neither is well correlated with surface area.  Instead, we determine the root mean square deviation of each structure's distance from the 
hypothetical ``ideal" flat-faced bubble: $F_{rms} = \langle (F -13.39)^2 \rangle$ and $e_{rms} = \langle (e_F -5.10)^2 \rangle$. 
For the F-K phases, the values of $e_{rms}$ are all similar, and do not distinguish the structures. Conversely, $F_{rms}$ is remarkably well correlated with the surface area for these structures, as shown in fig. \ref{fig:S_vs_rmsF}.
The Kelvin structure is likely to be very different to the structures derived from F-K phases; its value of $F_{rms}$ is low but its value of $e_{rms}$ is very high, and the straight line of  fig. \ref{fig:S_vs_rmsF} could be generalized into a fit to two variables $(e_{rms},F_{rms})$. In figure \ref{fig:S_vs_rmsF} we also show data for other F-K layered (Z, $\sigma$ and C15) and not layered (Bergman and S-M) structures generated by A. Kraynik.

\begin{figure}
\centerline{
\showfigures{\includegraphics[width=0.75\textwidth]{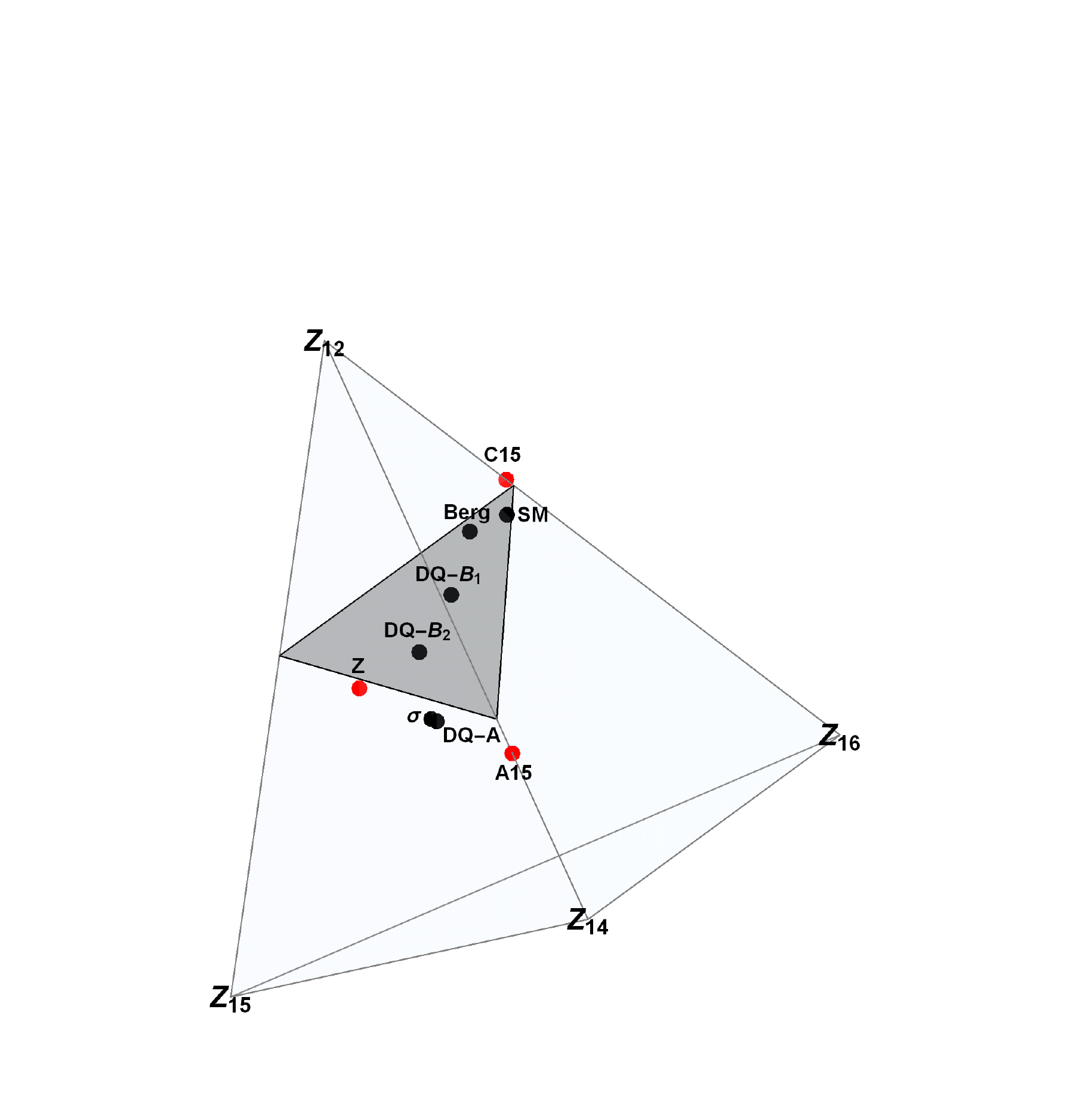} }
}
\caption{Comparison of topologies. Tetrahedrally-close-packed Frank-Kasper phases are plotted using barycentric coordinates according to their ratios of four possible Frank-Kasper canonical polyhedra: $Z_{12}$, $Z_{14}$, $Z_{15}$, and $Z_{16}$. The grey plane is the locus of structures having a mean coordination number equal to that of a Coxeter statistical honeycomb \cite{coxeter58}.
 A15 (dual to the Weaire-Phelan foam), C15, Z and $\sigma$ are standard F-K structures \cite{frankkasper}. Bergman (named Berg in the figure) and SM are closely-related but more complex TCP phases with 162 sites per unit cell and entangled disclination networks (SM is a one step decoration of the A15 phase). DQ-$\rm{A}$, DQ-$\rm{B}_1$ and DQ-$\rm{B}_2$ are the new quasiperiodic phases whose dual foams are described here.}
\label{fig:sullivantri}
\end{figure}

Bubbles organized following the F-K scheme, including W-P, dual to the A15 F-K phase, seem to provide interesting candidates to the Kelvin problem. In Fig. \ref{fig:sullivantri} (loosely inspired by Fig. 5 of \protect\cite{kusners96}), we display some classical TCP F-K phases, with barycentric coordinates relative to the main four Frank-Kasper canonical polyhedra ($Z_{12}$, $Z_{14}$, $Z_{15}$ and $Z_{16}$). The A15 phase has straight non-intersecting disclination lines in three directions (so with only $Z_{12}$ and $Z_{14}$ sites). The Z phase has a family of straight disclination lines in one direction orthogonal to planes containing disclination lines connected three by three (so with $Z_{12}$, $Z_{14}$ and $Z_{15}$). The Laves phases have disclination lines connected four by four, like a diamond network, represented here by C15 (with only $Z_{12}$ and $Z_{16}$ sites). Also shown on the figure are the quasiperiodic TCP structures whose dual foams have been numerically studied in this paper. It is striking that all this structures fall very close to a particular (grey) plane in the drawing, which is the locus of structures having a mean coordination number equal to that 
of a Coxeter statistical honeycomb \cite{coxeter58,coxeter61}.

\section{Conclusion}

This paper introduces numerically-derived quasiperiodic foams. We have been able to simulate finite structures, large enough to resemble their infinite theoretical counterpart, and found them stable. By simulating  periodic structures, which surface area is known, we find as expected that simulations of finite clusters slightly overestimate surface areas. Relative errors are small enough that we can confidently classify the structures on the scale of their simulated surface areas.
We argue that  $N=1500$ to 2000 bubbles is enough to obtain results barely sensitive to calculation details such as the position of the boundaries, and to the variation of $N$. 

While we failed to find  a quasicrystalline foam with a lower surface area than W-P,  the present study nonetheless contributes to the Kelvin problem. Enlarging the search for candidates from crystalline F-K phases to different quasicrystalline ones clarifies the role of a foam's main topological feature: the distribution of the number of faces $F$ per bubble and, more specifically, the rms deviation $F_{rms}$ of $F$ from 13.39. 

Our results are compatible with the conjecture that  $F_{rms}$ is a determinant of the surface area. Furthermore, we observe that  $F_{rms}$ increases with the complexity of the disclination network.
From this point of view, the W-P candidate is extremal among the Frank-Kasper duals, with only  $Z_{12}$ and $Z_{14}$ sites. As shown numerically here, the foams corresponding to structures containing $Z_{16}$ sites have larger surface areas than those having only $Z_{15}$ in addition to $Z_{12}$ and $Z_{14}$ sites.

Other parameters could play a role too. For instance, the density of the sites for bubble centres should probably be as homogeneous as possible. It is a property of the 2D quasicrystalline phase that it minimizes these density fluctuations. The Kelvin structure, which is not a F-K phase, also minimizes density fluctuations, but with a more complex disclination network and a high coordination number, $p=14$.

The structures analyzed here are not fully quasiperiodic, being still, by construction, periodic along one direction. A natural extension would be to study foam clusters constructed as a dual to an icosahedral quasicrystal, quasiperiodic in all three directions.  

Conversely, it is possible  (see the appendix of \cite{sadocmosseri16}) to build structures which are quasiperiodic in only one direction, with and without $Z_{16}$ sites, and periodic along the other two. In the representation of fig. \ref{fig:sullivantri}, these structures would fall very close to the A15 point, and can be seen as an A15 crystal interrupted by planar defects. They should therefore have  very close surface area values, presumably lower than those discussed in the present paper. It would be interesting to compare the two cases, with and without $Z_{16}$ sites, and check whether  the latter always increases the surface area of the corresponding foam.

Finally, dynamical considerations may prove interesting. When a foam is sheared beyond a certain deformation, called the ``yield strain",  it undergoes local topological transformations, respecting Plateau's laws, called ``T1s"  \cite{Cantat2013}. In a crystalline foam, the yield strain is well defined, in the sense that several T1s occur simultaneously, and is anisotropic, with a preferential yielding along crystalline directions. On the other hand, in an amorphous foam, the yield strain is isotropic, and less precisely defined, since precursory isolated  T1s occur below the actual yield strain. Quasiperiodic structures should belong to a new class. They are anisotropic, so their yield strain is probably anisotropic too. But most importantly,  quasiperiodic structures undergo  peculiar types of local rearrangement, called localized ``phasons" \cite{henley88}. These phasons are more complex, i.e. less localized, in the present dodecagonal quasicrystals than for icosahedral ones. It would be interesting to check whether these features could indeed be observed in quasiperiodic foams.

\subsection*{Acknowledgements}

We thank A. Kraynik for sharing data with us, and K. Brakke for providing and supporting the Surface Evolver software. SJC acknowledges funding from the MCSA-RISE project Matrixassay (ID: 644175).

\end{document}